# Acoustic Screens based on Sonic Crystals with high Diffusion properties

M.P. Peiró-Torres[1,4]; M.J. Parrilla Navarro[2]; M. Ferri[3]; J.M. Bravo[4]; J.V. Sánchez-Pérez[4]; J. Redondo[3]

(1) BECSA. Ciudad del Transporte II. C/ Grecia, 31, Castellón (Spain); mppeiro@becsa.es

(2) Universitat Politècnica de València. Paranimf 1, Grao de Gandia, Valencia (Spain); maparna@epsg.upv.es

(3) Universitat Politècnica de València. Instituto de Investigación para la Gestión Integrada de zonas Costeras, Paranimf 1, Grao de Gandia, Valencia (Spain); mferri@fis.upv.es; fredondo@upv.es

(4) Universitat Politècnica de València. Centro de Tecnologías Físicas, Acústica, Materiales y Astrofísica, División acústica. Camino de Vera s/n, Valencia (Spain); jobrapla@fis.upv.es; jusanc@fis.upv.es

**Abstract**

This article presents the use of advanced tools applied to the design of devices that can solve specific acoustic problems, improving the already existing devices based on classic technologies. Specifically, we have used two different configurations of a material called Sonic Crystals, which is formed by arrays of acoustic scatterers, to obtain acoustic screens with high diffusion properties by means of an optimization process. This design procedure has been carried out using a multiobjective evolutionary algorithm along to an acoustic simulation model developed with the numerical method called Finite Difference Time Domain (FDTD). The results obtained are discussed in terms of both the acoustic performance and the robustness of the devices achieved.

**Corresponding author:** Javier Redondo Pastor[3] (J. Redondo).

# 1. Introduction

Environmental noise can be defined as an unwanted or harmful outdoor sound created by human activities, and is one of the main environmental problems all over the world [1]. Among all types, traffic noise caused by cars and duty vehicles is one of the most important and annoying, making the greatest contribution to total noise pollution (around 90%) [2]. Traffic is behind the high noise levels experienced by European citizens, as according to the EU, noise levels above 55 dBA at night and 65 dBA during daylight hours should not be exceeded to ensure the comfort of citizens. However, EU-Eurostat states that 20% of EU citizens during the day and 30% at night suffer from higher noise levels. These high grades of exposure are linked with some health problems such as stress, sleep disturbance, fatigue, cardiovascular disorders or hearing loss [3,4].

Generally speaking, environmental noise can be mitigated (i) at the source, reducing the radiated sound power emitted by vehicles; (ii) during its propagation, reducing the noise level during its propagation from the source to the receiver or (iii) in the receiver, improving the isolation of the dwellings and preventing its transmission through the exterior walls. When the noise control is carried out in its propagation phase, the most used solution is the placement of acoustic barriers (AB) [5], which are located between the noise source and the receiver. Classical AB are generally made of continuous flat walls of different materials such as concrete, wood or methacrylate, and have to meet a certain number of standards in terms of their density and geometry to be acoustically effective [6]. The performance of AB can explained as follows (Fig.1(a)): noise is propagated from the source to the receiver following a straight line. AB are placed between them, and an important quantity of the noise energy is reflected specularly while other parts are diffracted from the edge of the barrier, transmitted through it or dissipated by the material that forms the barrier.

If we focus on the energy of specularly reflected noise, some unwanted problems can arise when placing AB to protect predetermined areas. Thus, sometimes the site where AB is located to acoustically protect a receiver can increase the noise level in other locations that also need protection. This situation is illustrated in Fig. 1(b) as an example, where the building A is protected by the AB A. However, the installation of another AB B to protect the building B can produce some reflections that increase the noise level received by the building A, reducing the effectiveness of the AB A [3,5]. This situation is quite common as show in Fig. 1(c), where the picture has been taken at one of the entrances to the city of Cádiz (Spain).

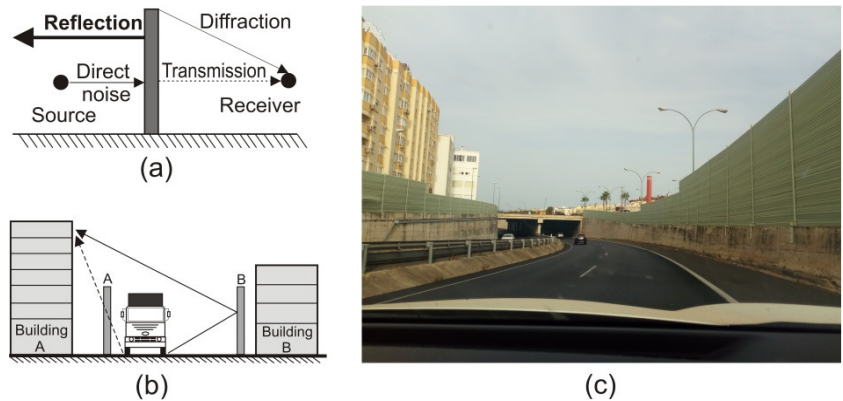

Fig. 1. (Color online) (a) Scheme of the acoustic performance of AB; (b) Scheme of the problems created by the specularly reflected noise; (c) A picture taken at one of the entrances to the city of Cádiz (Spain) to illustrate the described situation. (For interpretation of the references to color in this figure caption, the reader is referred to the web version of this article.)

In order to minimize these specular reflections, several solutions have been proposed, the most common of which are (i) the use of absorbent materials in AB; (ii) the construction of inclined AB, in such a way that the specularly reflected sound is diverted outside the areas to be protected; or (iii) the scattering of the reflected noise on AB, avoiding specular reflection [7]. However, the first two solutions present some problems related to their cost: the use of absorbent materials in AB could increase their price reducing their competitiveness, and the use of tilted AB can be even more expensive and their installation technically complicated for some sites.

Regarding the solution based on scattering reflected noise, some new proposals have been made in recent years. One of the most widely accepted is the use of new devices based on technologically advanced materials devoted to noise control. Sonic Crystals (SC), generally defined as heterogeneous materials formed by arrangements of acoustic scatterers embedded in air, is one of these materials [8,9]. There are many proposed applications for these materials, including acting as metamaterials [10,11], but in this work we will use two in particular. On one hand their use as AB [12,13], usually called Sonic Crystals Acoustic Screens (SCAS). In this application SC provide a new noise control mechanism by structuring the scatterers, which provides the existence of bandgaps, defined as ranges of frequency where the propagation of the waves is forbidden [14,15]. The existence of bandgaps is the result of the interference of waves due to a Bragg scattering within the SC. These new barriers present aesthetic and technological advantages thanks to their open structure and their versatility to be designed for specific noises, among others properties. However, SCAS also present the specularly reflection of noise, as classical AB.

On the other hand, the use of sound diffusers in room acoustics to increase the sound diffuseness is generally accepted for four decades ago, when Schroeder presented the first proposal of such devices [16]. Since then, several designs have been proposed [17,18,19,20] but again, SC seem good candidates to obtain high diffusion levels, even at low frequency range, using smaller device depths than in the case of conventional diffusers [21]. These technologically advanced devices, generally called Acoustic Sonic Crystal Diffusers (SCAD), as is the case with diffusers in general, do prevent specular reflection of noise.

In addition, in recent years it has been possible the increasing of the acoustic performance of some devices based on SC, as SCAS or SCAD, through the use of evolutionary algorithms. Specifically, an elitist Multiobjective Evolutionary Algorithm (MOEA), called ev-MOGA [22], has been used to go a step further in designing technologically advanced noise control devices based on SC, creating SCAS [23] and SCAD [24,25] with high acoustic control properties.

Following this research line, in this work we present the process of designing new devices based on SC that work simultaneously as SCAS and SCAD. To obtain this goal, we have varied the radii of the cylindrical scatterers that form a pre-selected SC module using a MOEA as a tool. Although the idea of designing devices with this double function -protecting against direct noise and avoiding specularly reflected noise- is not new [26] and it is generally carry out by adding a sound diffuser to classic AB [27,28] or designing classic AB with a corrugated side [3], our procedure is far away from these designs since we use advanced materials and new designing tools. These new devices will work fundamentally as AB but with a low level of specularly reflection, minimizing the disturbance that sometimes appears when AB are used to control transport noise. Hereafter we will refer to these new devices as SCASAD (Sonic Crystals Acoustics Screens and Diffusers).

The paper is organized as follows: in Sec. 2 we describe the optimization process, explaining both the optimization tool and the simulation model used. The results obtained in the optimization process are analyzed and discussed in Sec. 3. The last section, Sec. 4, contains the closing remarks, where the main conclusions are summarized.

## 2. Theoretical considerations

### 2.1. Description of the optimization process.

In this section we briefly explain the main characteristics of the MOEA used in this work as well as the optimization procedure carried out. There are certain types of optimization problems in which is necessary to achieve solutions that satisfy several objectives simultaneously. Obviously, the natural tendency is to search the best solution for each one of the considered objectives. However, if the objectives are in conflict, usually an improvement in one of them means a worsening in others, and this means that there is not a single optimal solution. These kinds of problems, where several conflicting objectives have to be simultaneously optimized are known in the literature as multiobjective optimization problems, and they may be solved using MOEA [29]. A general basic multiobjective problem can be formulated as follows:

$$\text{Eq. (1): min } J(\theta) = \min [J_1(\theta), J_2(\theta), \ldots, J_s(\theta)],$$

subject to $\theta_{li} \leq \theta_i \leq \theta_{ui}$ ($1 \leq i \leq L$),

where $J_i(\theta)$, $i \in B := [1 \ldots s]$ are the objectives to be minimized, $\theta$ is a solution inside the $L$-dimensional solution space $D \subseteq R^L$, and $\theta_{li}$ and $\theta_{ui}$ are the lower and the upper constraints that defined the solution space $D$.

The general way to solve such problems using MOEA is the localization of a set of infinite optimal solutions in the objective space, which is mapped as the Pareto front. This front shows the best individuals, in some sense, obtained in the optimization process and classified according to the values achieved in the functions to be optimized. The basic concept to obtain the Pareto set is known as Pareto dominance, which is defined as follows: a solution $\theta^1$ dominates another solution $\theta^2$, denoted by $\theta^1 \prec \theta^2$, if $\forall i \in B$, $J_i(\theta^1) \leq J_i(\theta^2) \wedge \exists k \in B: J_k(\theta^1) < J_k(\theta^2)$. The Pareto set $\Theta_P$ is composed by all the non-dominated solutions, and the associated Pareto front is denoted as $J(\Theta_P)$. Due to the difficulties appeared in real problems to get the exact Pareto front, we have used here an elitist multi-objective evolutionary algorithm based on the concept of e-dominance [30] named ev-MOGA [22]. A complete explanation of the foundations and functioning of this algorithm as well as its applications in the field of SC can be found in references [22,23,24].

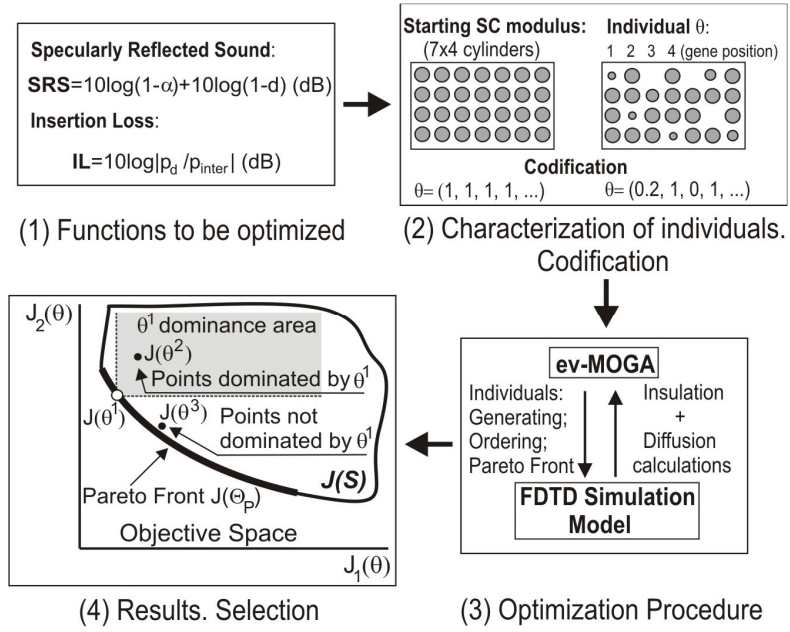

Fig. 2. Scheme of the steps followed in the optimization process

An outline of the optimization procedure is shown in Fig. 2. First (step 1), it is necessary to define the functions to be optimized, generally referred to as optimization objectives or cost functions. In our case, we want to design devices with high levels of acoustic insulation and diffusion. This means a bi-objective optimization procedure and we have to carefully define two cost functions, related to these properties, to characterize the effectiveness of our devices. Its definition must take into account the characteristics of the ev-MOGA algorithm, which works minimizing cost functions.

The first cost function we have chosen, related to the acoustic insulation capabilities of different individuals, is related to the Insertion Loss (IL) index, defined as the difference in sound pressure at a point or area with and without the sample. Note that the goal is to achieve a high level of acoustic insulation and therefore, in the optimization process we will work minimizing –IL. Thus, for a solution (individual) θ:

$$\text{Eq. (2):} \quad J_{\text{-IL}}(\theta) = -10 \log \left| \frac{P_d}{P_{inter}} \right| \text{ (dB)}$$

where $p_d$ is the direct acoustic pressure (without device), and $p_{inter}$ is the acoustic pressure interfered (with device), both calculated at the same point or area.

The second cost function concerns the diffusion properties of the individuals. To characterize this acoustic capability we have defined a new index called Specular Reflection Sound (SRS). For an individual θ is defined as:

$$\text{Eq. (3): } J_{SRS}(\theta) = 10\log(1-\alpha) + 10\log(1-d) \text{ (dB)}$$

where $\alpha$ is the absorption coefficient and d is the diffusion coefficient for each individual.

These two cost functions determine the performance of individuals as both SCAS and as SCAD in the predetermined range of frequencies stablished by us. In this work we have selected a range of frequencies formed by the octaves bands whose central frequencies are 500 Hz, 1000 Hz and 2000 Hz, i.e. a range of frequencies from 355 Hz to 2828 Hz. The reason for this selection is related with the nature of the noise that our devices will deal with, which is given by the normalized spectral traffic noise defined in the norm UNE 1793-3:1998, where critical frequencies are covered by our selected range.

Once the cost functions have been defined, the next step of the optimization procedure (step 2) is twofold: (i) the characterization of the shape of the individuals –including the initial population with which the optimization process begin- in such a way that the population will be formed by a variable set of individuals, all of them based on a predetermined SC module, and (ii) their codification. In this work we have selected a module formed by 28 cylindrical rigid scatterers arranged in 4 rows. The reason for this choice is related to the characteristics of the SCAS and SCAD designed and/or optimized up to now: SCAS are usually formed by 3 or 4 rows [12,13] and, at the same time, SCAD are formed by 4 rows [24]. Taking these results into account, an optimized SCASAD should consist of at least 7-8 rows, adding the necessary rows for an optimal performance as SCAS and SCAD. However, our design proposal aims to force the acoustic performance of the CS to produce a very compact device made up by the fewest number of rows, set by us at 4, in order to obtain an occupancy similar to that of the classic ABs at road shoulders. In addition, the number of scatterers in each row ensures a reasonable genetic variation of the population taking into account the tool selected to obtain new individuals from the initial population, as we explain below. This initial module does not have a high performance as either SCAS or SCAD, due to the low number of rows that compose it, and its insulation and diffusion properties will be greatly improved in the optimization process to be carried out.

On the other hand, in order to provide enough genetic variation to the initial population necessary to create new individuals with a high variability in the values of their cost functions, we have used as a tool the variation of the radii of the cylindrical scatterers of the individuals formed from the module previously defined (7x4 cylinders). To characterize each individual of the population it is necessary to establish a gene codification, encoding each one of them by means of a set of genes that represents the set of the 28 (7x4) normalized cylinders radii. Each radius can take any value from 0 to 0.9. If the value is 0, the cylinder does not exist and, if the value is 0.9, the cylinder has almost the maximum possible radius, which is equal to the half lattice constant. In this way, any individual $\theta$ can be represented by a genotype given by a vector of length 28 elements, varying each one from 0 to 0.9. Two examples of the genetic coding are shown in Fig. 2.

Once the cost functions and the codification of individuals have been defined, the optimization procedure can be initiated. This process works using together ev-MOGA and an acoustic simulation model developed by us, which will be presented in next section. ev-MOGA leads the process (i) generating new individuals by mixing, following the rules of genetics including mutations, the genotypes of the individuals from the initial population generated by us; (ii) ordering and representing the different individuals in the objective space according to the values of each of the defined cost functions and (iii) stablishing the Pareto Front in the objective space. On the other hand, the simulation model evaluates the acoustic performance of each individual generated by ev-MOGA, calculating the values of its cost functions (-IL and SRS) and providing them to ev-MOGA. Finally, the optimization procedure ends when an optimal solution belonging to the Pareto Front obtained is selected according to designer preferences.

**2.2. Simulation model: Finite Difference Time Domain**

To acoustically characterize the different individuals obtained in the optimization process, we have developed a simulation model based on the numerical technique called Finite Difference Time Domain (FDTD). This model works together with ev-MOGA and performs the necessary calculations to obtain the values, for each individual, of the previously defined cost functions. FDTD is often used in acoustic simulations of different devices. In particular, it has been already used successfully to quantify the acoustic performance of SC in some optimization processes, working together with ev-MOGA [24]. Further details about the characteristics of this numerical setup can be found in reference [31].

The model developed specifically for this paper is shown in Fig. 3. The rectangular calculation domain is formed by two parallel lines with periodic boundary conditions in order to simulate a semi-infinite SC. Furthermore, to avoid unwanted reflections, a Perfectly Matched Layer (PML) is located at the right of the domain.

With these boundary conditions, the numerical scheme is excited by a line source placed at the left hand side of the integration area (see Fig 3(a)). As FDTD works in the time domain it is extremely important to use excitations signals as short in time as possible in order to minimize the computational cost. In this work we have used a Dirac delta filtered with the normalized traffic noise spectra defined in the UNE 1793-3:1998 norm. Part of this generated signal is transmitted through the device and another part is reflected to the left. The insulation performance of each individual, given by the -IL cost function, is calculated behind the SC, on the right area of the model (measurement area in Fig. 3). To do that, we have obtained the acoustic pressure every 0.02 m in this area, with and without the sample, to obtain the -IL value at each point in 1/3 octave band for the selected range. Then a spatial average has been carried out to obtain a single -IL value for each individual.

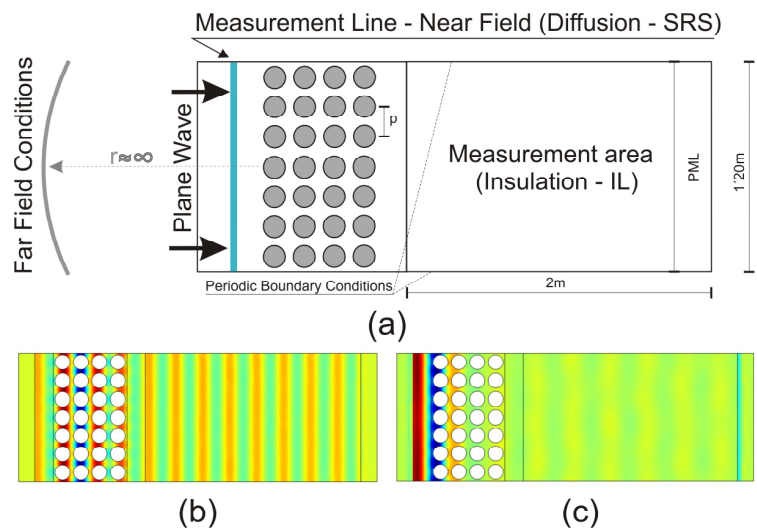

Fig. 3. (Color online) (a) Scheme of the simulation model, based on FDTD, used to acoustically characterize the individuals generated by ev-MOGA; (b) and (c) Examples of acoustic pressure fields, in Pa, for frequencies outside (1370 Hz) and inside (1000 Hz) the bandgap respectively. (For interpretation of the references to color in this figure caption, the reader is referred to the web version of this article.)

On the other hand, to calculate the SRS index we need to estimate both the incident and the reflected sound. For doing so, the total acoustic pressure is recorded every 0.01 m in 1/3 octave bands along the entire vertical blue line shown on the left in Fig. 3 (measurement line). By comparison between the cases with and without the sample, the incident and reflected sound are estimated. From these data we can obtain the absorption coefficient (alfa). Finally, from the reflected sound data, we have carried out a near field to far field transformation in order to estimate the value of the diffusion coefficient, d, in free field conditions according to the standard ISO 17497-2-2012

According to the characteristics of the numerical model explained, the optimization process is developed only for the normal incidence of the wave on the SC.

**3. Results and discussion**

To obtain high performance devices that act simultaneously as SCAS and as SCAD we have used in this work the combination of ev-MOGA and FDTD, as we have commented above. One of the main problems of this optimization procedure is that the joint use of both algorithms implies large computational cost. In our case, the FDTD simulation for each device takes about 240 s on an Intel Core i7-3632QM 2.20 GHz (Santa Clara, CA). To calculate the total runtime, it is necessary to take into account that the total number of calculations in the optimization process is estimated as the number of new individuals plus the number of individuals in the initial population. Once the Pareto set is obtained for each generation, it is used as part of the initial population for the next optimization. In the process of optimization developed in this work, the initial population is formed by 2000 individuals, and in each generation 8 new individuals are added. Under these conditions, the total execution time of the entire process is 7 days, considering 1000 generations.

The selected configuration is formed by the initial module defined in section 2.1 but arranging the cylindrical scatterers using only one lattice constant. That means that the existing bandgaps in the region of interest are due to only one periodicity. Specifically, and taking into account the normalized spectral traffic noise defined in the norm UNE 1793-3:1998, we have set the value of the considered single lattice constant in p=0.17 m, which corresponds to a first bandgap centered at 1000 Hz for an incidence of 0º on the SC, the most critical frequency of the normalized spectral traffic noise. Additionally, the following bandgaps (second and third) would be located at 2000 Hz and 3000 Hz respectively, within the frequency range of interest. With these geometrical conditions, the objective in this optimization process is to design a SCASAD device with high performance, simultaneously, as SCAS and SCAD around the same global target frequency, 1000 Hz.

The results of the optimization are shown in Fig. 4(a), where the objective space is represented. The black dots represent the individuals of the initial population according to their single values of both cost functions considered, -IL and SRS (abscissa and ordinate axes respectively) calculated as shown in Section 2.2. The individual belonging to the initial population, which is formed by cylinders of equal radius corresponding to a filling fraction of 75%, (r=0.08 m), is represented in the Figure by a blue diamond and is called by us "reference individual". The position of this non-optimized individual in the objective space serves as a reference for the improvement achieved in the optimization process. The Pareto Front is formed in the Fig. 4(a) by the individuals marked as red squares. Among all the individuals that form the Pareto Front, we have selected as designers the individual marked with a green square due to its balanced values of both cost functions, and we have named it "selected individual". Fig. 4(b) shows the individuals considered, the reference (top) and the selected one (below).

The acoustic performance of both individuals (the reference in continuous blue line and the one selected in dashed green line) can be seen in Fig. 4(c) as a function of frequency, where the range of interest of the study is also indicated. Note that the IL and –SRS indexes, instead –IL and SRS, are represented here for better understanding. The insulation (IL) spectra for both individuals are shown at the top of Fig. 4(c), where the higher global performance trend as SCAS of the reference individual compared with the selected one (15,7 dB versus 12 dB in Fig. 4(a)) can be checked. In addition, the first bandgap of the arrangement at 1000 Hz, corresponding to the considered lattice constant, can be easily observed for both individuals, wider in the case of the reference individual and smaller for the selected one. On the other hand, the diffusion properties of both individuals, given here by the -SRS index, are shown in the center of Fig. 4(c). As can be seen, the -SRS values are generally higher for the selected individual and lower for the reference, confirming the trend shown in Fig. 4(a) (3 dB versus 0.8 dB).

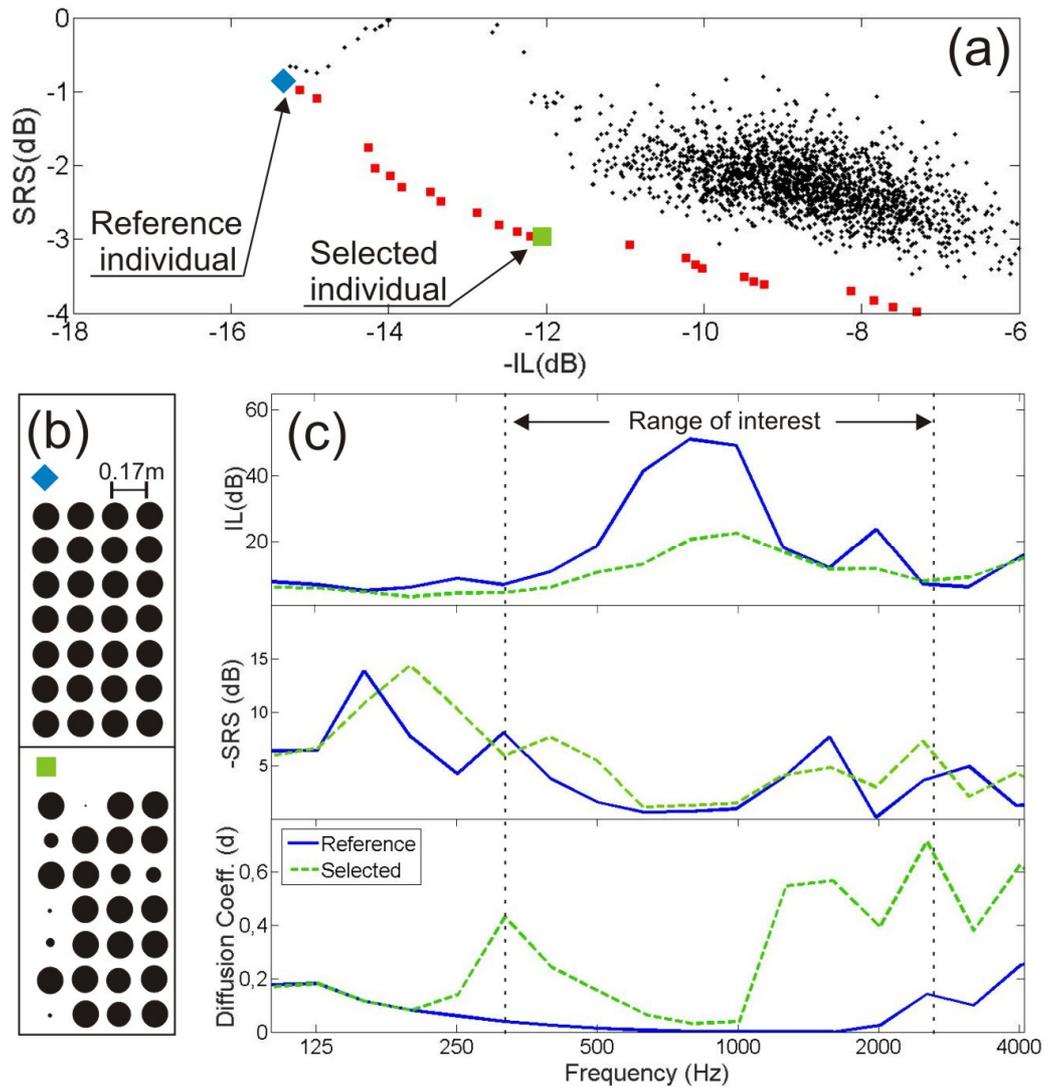

Fig. 4: (Color online) Optimization results for the analyzed case. (a) Objective space where the initial population, the Pareto Front and both the selected and the reference individuals are remarked; (b) Analyzed devices; (c) Acoustic performance of both individuals, reference (blue continuous line) and selected (green dashed line). IL, -SRS and d spectra are shown in the target frequency range. The spectra are calculated in 1/3 octave bands (For interpretation of the references to color in this figure caption, the reader is referred to the web version of this article.)

Interesting conclusions can be drawn from the results of this optimization process. Firstly, the increase in diffusion properties in the optimization is achieved at the expense of loss of attenuation capability. Thus, in the case of the selected individual, an increase of 2.1 dB in the SRS cost function implies a loss of 3.7 dB in the -IL index with respect to the corresponding values obtained by the reference individual (see Fig. 4(a)). Second, it seems that the increase of the diffusion capabilities of the optimized individuals is quite small compared with the SRS values that the initial population has. However, this conclusion seems a consequence of the selected cost function (SRS). Indeed, if in the frequency range considered we analyze the value of the diffusion coefficient d, used to measure the diffusion capability according to current standards and represented at the bottom of Fig. 4(c), we can conclude that the mean value of the diffusion coefficient of the selected individual compared to the reference increases considerably (0.3 versus 0.02). Finally, analyzing the -SRS and IL spectra represented in Fig. 4(c), it can be concluded that the higher the insulation value, the lower the -SRS value. This fact can be seen for the two individuals analyzed, although it is more remarkable in the case of the selected individual: around the bandgap frequency (1000 Hz), where the insulation values are maximum, minimum values of -SRS and d appear. This result is of great importance for the design of SC-based devices: it is not possible to create SC with high performance as an insulator and as a diffuser in the same frequency target, since a high attenuation implies low diffusion. The explanation of this fact could be related to the small number of SC rows considered. We think that we have pushed SC to the limit of their acoustic performance, demanding that they work as insulators and diffusers with only 4 rows, a very small number. Perhaps with more rows their acoustic performance could be increased. But the initial requirements force us to maintain that number of rows so that these devices are competitive with respect to the existing ones.

### 3.2. Study of the robustness of the selected devices

Another interesting parameter used to help decision maker to choose the most appropriate individual in the optimization process is the robustness of the selected devices. This concept has been previously introduced by some of us, and is defined as the degree to which the values of cost functions are affected by small changes in the values of the parameters that vary in the optimization process [24]. In our case, we have studied the robustness of the devices related to the variation of the cylinder radius that may appear due to possible errors in the manufacturing process. The low robustness of an individual means that it may not be the right choice, as some small unwanted and uncontrolled variations in cylinder radii can result in a significant reduction in the acoustic performance of devices.

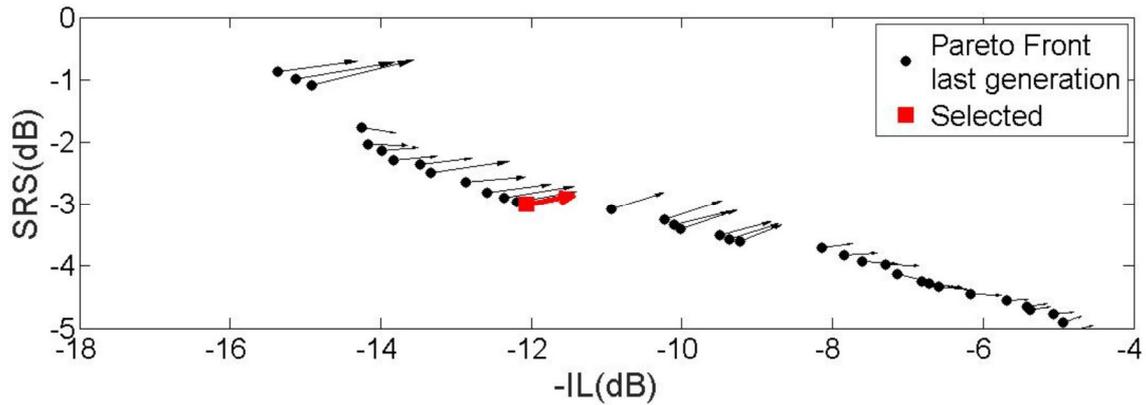

Fig. 5. (Color online) Pareto front with the robustness vectors of the radii the analyzed case. Both selected individuals are represented by a red square, and the particular robustness of both individuals is shown by a thick red vector. (For interpretation of the references to color in this figure caption, the reader is referred to the web version of this article.)

The robustness of individuals is represented by a vector that provides information on each individual according to the following rules: (i) the size of the vector indicates how robust an individual is according to the following rule of thumb: the larger the size of the vector, the less robust it is; (ii) Also, the size of vector components along the axes that represent the cost functions indicates how robust the individual is relative to each of them. To obtain the robustness vectors, each individual of the Pareto Front has been recalculated 200 times producing small random variations in the radii of the cylinders that form it. To simulate some defects in this manufacturing process, we have modified the radius of all cylinders by 5%. In doing so, we obtain in the objectives space a cloud of points around each initial Pareto point. This cloud is averaged at a single point and, finally, the robustness vector, whose origin is the point of Pareto considered and its end is this point average of the modified individuals, is plotted.

Fig. 5 shows the robustness vectors of the individuals forming the Pareto Front, including the selected ones. It can be seen in the Figure that the robustness is greater in the Pareto points with high SRS and low IL, and lower when Pareto individuals present low SRS and high IL. Another interesting conclusion that can be drawn from Fig. 5 is that the horizontal components of the vectors robustness (components according to IL) are generally greater than the components according to the vertical axis (components according to SRS). This fact indicates that the IL variable is less robust than the SRS variable. Moreover, from Fig. 5 it can be concluded that all the robustness vectors represented indicate that any variation in the radii of the scatterers would produce individuals with lower acoustic performance than those belonging to the Pareto front. This is a good indicator that the optimization process has been carried out to the end.

## 4. Conclusions

In this work we have used a specific Multiobjective evolutionary algorithm, called ev-MOGA, together with a simulation acoustic model based on the numerical technique called Finite Difference Time Domain (FDTD) in a bi-objective optimization process. With these tools we have designed technologically advanced devices based on Sonic Crystals. Specifically, we have solved an environmental noise problem related to the performance of classical noise barriers. These barriers, generally formed by straight walls, reflect noise specularly, so that these reflections can cause nuisance on the opposite side of the place where the barriers are located. To solve this problem we have carried out an optimization with two cost functions related to the insulation and the diffusion of the devices, represented by the IL and SRS indexes respectively. The starting point of our designs is the use of a minimum number of rows of the SC, four, to obtain a new acoustic screen with high diffusion properties and the lowest possible thickness so that it can be installed on the roadside shoulders without space problems. Even with this important restriction, the results obtained are successful, in terms of both acoustic performance and robustness.

Although the existence of some problems related to the acoustic behaviour of sonic crystals, in particular the fact that the frequency ranges with maximum attenuation (bandgap) correspond to the minimum diffusion, makes it difficult to obtain devices with the proposed dual function of attenuator and diffuser, we have been able to obtain advanced devices based on SC to control noise. The resultant devices have been called Sonic Crystals Acoustic Screens and Diffusers (SCASAD) by us, and provide a high technological design process to solve an environmental problem with the help of new materials and tools.

## Acknowledgments

This work was partially supported by the Spanish "Ministerio de Economia y Competitividad" under the project TEC2015-68076-R.

*'Declarations of interest: none'*